\documentclass{aastex}
\usepackage{graphicx}
\usepackage{color}

\usepackage{ulem}

\begin{document}

\title{Counter-rotation in relativistic magnetohydrodynamic jets}
\shorttitle{Counter-rotation in relativistic MHD jets}
\shortauthors{Cayatte et al.}

\author{V. Cayatte\altaffilmark{1}}

\author{N. Vlahakis\altaffilmark{2}}

\author{T. Matsakos\altaffilmark{3, 4}}

\author{J.J.G. Lima\altaffilmark{5,6}}

\author{K. Tsinganos\altaffilmark{2,7}}

\author{C. Sauty\altaffilmark{1}}

\email{veronique.cayatte@obspm.fr}

\altaffiltext{1}{Laboratoire Univers et Th\'eories, Observatoire de Paris, UMR 8102 du
 CNRS, Universit\'e Paris Diderot, F-92190 Meudon,  France}

\altaffiltext{2}
{Department of Astrophysics, Astronomy and Mechanics,
Faculty of Physics, University of Athens, 15784 Zografos, Athens, Greece}

\altaffiltext{3}{Department of Astronomy \& Astrophysics, The University of Chicago, Chicago, IL 60637, USA}

\altaffiltext{4}{LERMA, Observatoire de Paris, Universit\'e Pierre et Marie Curie,
  Ecole Normale Sup\'erieure, Universit\'e Cergy-Pontoise, CNRS, France}

\altaffiltext{5}{Centro de Astrof\'{\i}sica, Universidade do Porto, Rua das Estrelas,
 4150-762 Porto, Portugal}

\altaffiltext{6}{Departamento de F\'{\i}sica e Astronomia, Faculdade de
 Ci\^{e}ncias, Universidade Porto, Rua do Campo Alegre, 687, 4169-007 Porto,
 Portugal}

\altaffiltext{7}{National Observatory of Athens, Lofos Nymphon, Thission 11810, Athens, Greece}

\begin{abstract}
Young stellar object observations suggest that some jets rotate in the opposite
direction with respect to their disk. 
In a recent study, \cite{Sautyetal12} have shown that this does not contradict
the magnetocentrifugal mechanism that is believed to launch such outflows.  
Signatures of motions transverse to the jet axis  and in opposite directions have recently been measured in 
M87 (\citeauthor{Meyeretal13}, \citeyear{Meyeretal13}). One possible
interpretation of this motion is the one of counter rotating knots.
Here, we extend our previous analytical derivation of counter-rotation to relativistic jets, demonstrating
that counter-rotation can indeed take place under rather general conditions.
We show that both the magnetic field and a non-negligible enthalpy are necessary 
at the origin of counter-rotating outflows,
and that the effect is associated with a transfer of energy flux from the matter 
to the electromagnetic field.
This can be realized in three cases : if a decreasing enthalpy causes an increase of the Poynting flux,
if the flow decelerates, or, if strong gradients of the magnetic field are present.
An illustration of the involved mechanism is given by an example of relativistic MHD jet simulation. 

\end{abstract}


\section{Introduction}

In a previous Letter \citep{Sautyetal12} we have established that counter rotation 
in jets from young stars could be a natural consequence of the MHD equations ruling the  
plasma. We have shown that 
deceleration of the jet or shocks can induce counter rotation. We have verified it analytically and numerically.
In young stars, it is possible to 
measure observationally the rotation speed of the jet. Counter-rotation has been observed 
in some cases but remains under debate. However, in the light of our criterion, it is clear that it does not 
contradict the magnetorotational launching of the jet.  

In the context of relativistic outflows such as AGN jets, the criterion, if it can be extended, may induce 
crucial observational consequences. Those jets are well known to be 
magnetically launched as well. 
Recent measurements of the polarization and the VLBI Faraday rotation (e.g. \citealp{Mahmudetal13, 
Algabaetal13}) have confirmed the helicoidal nature of the jet magnetic field which  supports the idea of 
magnetic launching. Faraday rotation measures provide a direct evidence of the magnetic field structure 
within the jet \citep{Gabuzda03} and allow to confront observations with simulations
\citep{BroderickMcKinney10}.
Besides, \cite{Meyeretal13} have measured  
transverse proper motions of knots in the jet of M87. They infer that knots A and C 
seem to have opposite velocities transverse to the jet 
which they interpret as counter rotating shocks. They claim that this is consistent with the model of quad relativistic MHD shocks of  
\cite{Nakamuraetal10}. This model was used to interpret the helical magnetic structure inferred from polarization measurements for the same knots by  
\cite{Algabaetal13}. It is worth to note that these observations are rather difficult and need a very long time survey.

\cite{KVKB09} have shown numerically that counter-rotation appears in their simulations (hot jet of model B2H) as the result of transfer of angular 
momentum from the fluid to the magnetic component. \cite{Nakamuraetal10} have also shown that the complex structure of the quad relativistic MHD 
shock model exhibits a reverse shock that flows upstream, rotating in the direction opposite to the forward shock. 

We show in this Letter a straightforward extension in the relativistic regime of our criterion for counter rotation in young stars 
which can also be interpreted in terms of the flow energetics. 
This criterion applies to shock models as well as simulations where the angular momentum and isorotation frequency are conserved. 
This does not necessarily means it applies to quad shock models. The compatibility with the Riemann problem solution still needs to be checked. However, the present criterion involved a simpler geometry and does not rely on the presence of kink instabilities or precession of  the jet axis.  

Then, even though rotation measurements in relativistic sources are still out of reach, 
counter-rotation may have strong observational signatures related to the magnetic field structures. In fact,  
precise measurements of the magnetic field gradients would be needed to get constraints on the jet 
dynamics and rotation.

\section{Steady axisymmetric relativistic MHD outflows}

\subsection{Integrals of motion in Kerr metrics}

Under the assumption of steady-state and axisymmetry the equations of 
general relativistic magnetohydrodynamics can be partially integrated
to yield several field/streamline constants (e.g., \citealp{Beskinbook}),
including the magnetic field angular velocity
\begin{equation}
  \Omega = \omega + \frac{h}{\varpi}\left(
    V_\varphi- \frac{V_{\rm p}}{B_{\rm p}}B_\varphi
  \right)\,,
  \label{Omega}
\end{equation}
the total angular momentum flux to mass flux ratio
\begin{equation}
  L = 
    \xi\gamma \varpi V_\varphi 
    - \frac{\varpi B_\varphi B_{\rm p}}{4\pi\gamma\rho_0 V_{\rm p}} \,,
  \label{L}
\end{equation}
and the total energy flux to mass flux ratio
\begin{equation}
  \mu c^2 = h\xi\gamma c^2+ \xi\gamma\omega\varpi V_\varphi
    - \frac{\varpi\Omega B_\varphi B_{\rm p}}{4\pi\gamma\rho_0 V_{\rm p}} \,.
  \label{mu}
\end{equation}
Here $h$ is the lapse function, 
$\varpi(=\sqrt{g_{\varphi\varphi}})$ is the cylindrical radius in Kerr geometry, 
$\omega$ is the angular velocity of zero angular momentum observers (ZAMO),
$B_{\rm p}, B_\varphi, V_{\rm p}, V_\varphi$ 
denote poloidal and toroidal components of the magnetic field 
and bulk flow speed as seen by ZAMO, $\gamma$ is the bulk Lorentz factor, and
$(\rho_0, \xi c^2)$ are the mass density and specific enthalpy 
as measured in the frame comoving with the outflow.

The system of the previous three equations can be solved for 
$(\gamma, B_\varphi, \gamma V_\varphi)$ as it is usually done. Equivalently  
we may use instead $(h\gamma\xi c^2, B_\varphi, \gamma \xi V_\varphi)$. 
This gives the following expressions in the observer's frame for the 
matter part of the energy flux to mass flux ratio
\begin{equation}
 h\gamma\xi c^2= \frac{M^2(\mu c^2-L\omega) - h^2(\mu c^2-L\Omega)}
{M^2 - h^2 + {\varpi^2(\Omega-\omega)^2}/{c^2}}\,,
\label{hgamma}
\end{equation}
and the toroidal proper component of the proper specific momentum
\begin{equation}
 \xi \gamma V_\varphi = \frac{L}{\varpi}\frac{M^2
    - {\varpi^2(\Omega-\omega)}(\mu c^2-L\Omega)/{Lc^2}}
   {M^2 - h^2 + {\varpi^2(\Omega-\omega)^2}/{c^2}}\,,
\label{vphi}
\end{equation}
in terms of the square of the ``Alfv\'enic'' Mach number
\begin{equation}\label{M2}
M^2={4\pi\xi\rho_0} h^2\gamma^2 V_{\rm p}^2 / B_{\rm p}^2 \,.
\end{equation}

The numerators and denominators of Eqs.~(\ref{hgamma})--(\ref{vphi}) vanish
at the Alfv\'en surface (subscript $\star$)
where the values of $M^2$, $\varpi$ are given by
\begin{equation}\label{M2star}
  \frac{M_\star^2}{h_\star^2} =\frac{\mu c^2-L\Omega}{\mu c^2-L\omega_\star} \,,
\quad
 \varpi_\star^2 = \frac{Lh_\star^2 c^2}
{(\Omega-\omega_\star)(\mu c^2-L\omega_\star)}\,.
\end{equation}

\subsection{Generic criterion in Kerr metrics for counter-rotation}

Assuming that the flow remains everywhere super-Alfv\'enic after crossing the
corresponding critical surface, the denominator of Eq.~(\ref{vphi}) is
always positive.
The sign of the toroidal velocity is then given by the sign of the numerator,
and thus we get that it is negative if,
\begin{equation}\label{Vphidem1}
 M^2< \varpi^2 (\Omega-\omega) \frac{\mu c^2-L\Omega}{Lc^2} \,.
\end{equation}

Along a given flux tube, we can write the magnetic and mass flux conservation as
\begin{equation}\label{Bflux}
  B_{\rm p}\delta S= B_{\star}\delta S_{\star} = {\rm constant}
\end{equation}
and
\begin{equation}\label{massflux}
h  \gamma \rho_0 V_{\rm p}\delta S= h_{\star}  \gamma_{\star}  \rho_{0\star} V_{\star}\delta S_{\star}= {\rm constant}
  \,.
\end{equation}
Note that $\delta S$ is the surface element perpendicular to the poloidal
velocity.

Using the definition of $M^2$ and the last two conservation laws 
following the procedure of the previous paper, we show that 
the condition~(\ref{Vphidem1}) is equivalent to,
\begin{equation}\label{Vphidem2}
\frac
  {\displaystyle 4\pi \rho_{0\star} h_\star \gamma_\star V_{\star} \frac{\displaystyle \delta S_{\star}}
  {\displaystyle \delta S} }
  {\displaystyle B_{\star}^2 \frac{\displaystyle \delta S_{\star}^2}
  {\displaystyle\delta S^2 }} h \xi \gamma V_{\rm p}
  < \varpi^2 (\Omega-\omega) \frac{\mu c^2-L\Omega}{Lc^2}
   \,.
\end{equation}

Using Eqs.~(\ref{M2star}) and (\ref{M2}) at the Alfv\'en surface,
the above expression simplifies further,
\begin{equation}
  \label{Vphidem31}
 \frac
 {\displaystyle h \xi \gamma V_{\rm p} \delta S}
{\displaystyle h_\star \xi_\star \gamma_\star V_{\star} \delta S_{\star}}
 < \frac
  {\displaystyle \varpi^2 (\Omega-\omega)}
  {\displaystyle \varpi_\star^2 (\Omega-\omega_\star)}
\,.
\end{equation}
Thus, the last inequality determines if the toroidal velocity is negative.

If the flow remains super-Alfv\'enic, reversal of the rotation takes place when
the proper velocity times the enthalpy drops below some threshold value.
The flow can decelerate or lose enthalpy  either because it expands and this causes adiabatic
cooling, or because its kinetic energy drops due to some other mechanism such as
radiative losses.
Note that we used in Eq. (\ref{Vphidem2}) energy conservation. However the criterion in  Eq. (\ref{Vphidem31}) remains valid if  the energy downstream, 
$\mu_{\rm local}$,  is lower than the energy at the Alfv\'{e}n surface because in that case,
$$
\varpi^2 (\Omega-\omega) \frac{\mu_{\rm local} c^2-L\Omega}{Lc^2}<\varpi^2 (\Omega-\omega) \frac{\mu_\star c^2-L\Omega}{Lc^2}
\,.
$$
Rotation reverses also if the threshold value increases and gets larger than the jet poloidal velocity.

In the sub-Alfv\'enic regime, close to the Black Hole, counter rotation may occur as well though the reverse inequality holds.
Above some threshold value of the energy, the rotation velocity reverses. In other words very close to the
central object, counter-rotation may occur if there is a steep acceleration followed by a mild deceleration 
before the Alfv\'en point. This is the case of the solution presented for RY Tau in \cite{Sautyetal11}. We 
have explored a generalization of this solution in Kerr metrics and we suspect that rotation also reverses in 
the sub-Alfv\'enic regime there. 

In the far asymptotic regime where the flow is super-Aflv\'enic and special relativistic 
($h \simeq 1$ and $\omega \ll \Omega$), the flow rotation reverses if,
\begin{equation}
  \label{Vphidem3}
 \frac
 {\displaystyle \xi \gamma V_{\rm p} \delta S}
{\displaystyle \xi_\star \gamma_\star V_{\star} \delta S_{\star}}
 < \frac {\displaystyle \varpi^2} {\displaystyle \varpi_\star^2 } \,.
\end{equation}

If the flow is sufficently smooth with $\delta S \propto \varpi^2$ and with ultrarelativistic 
velocities ($V_{\rm p}$ and $V_{\star} \simeq c$), the inequality simplifies to,

\begin{equation}
\xi \gamma < \xi_\star \gamma_\star \, .
\end{equation}

For an accelerated flow, this inequality implies that $\xi = 1\, + \, w/c^2$, where $w$ is the part 
of the specific enthalpy without the rest-mass density, should decrease after the Alfven surface.

\subsection{Criterion in terms of the flow energetics}
\label{sec:energetics}

In the following section, we restrict our discussion to a flow with constant energy flux.
The hydrodynamic and electromagnetic contributions in the 
angular momentum and energy fluxes can be easily identified by
inspecting Eqs.~(\ref{L}) and (\ref{mu}).
We can write $L=L_\mathrm{HYD} + L_\mathrm{MAG}$
and $\mu=\mu_\mathrm{HYD} + \mu_\mathrm{MAG}$, where
$L_\mathrm{HYD}=\xi\gamma \varpi V_\varphi$,
$L_\mathrm{MAG}=- {\varpi B_\varphi B_{\rm p}}/({4\pi\gamma\rho_0 V_{\rm p}})$,
$\mu_\mathrm{HYD}=h \xi\gamma +\xi \gamma \omega \varpi V_\varphi/c^2$,
$\mu_\mathrm{MAG}=L_\mathrm{MAG}\Omega/c^2$.

Near the regime where the toroidal speed changes sign from positive to negative,
$L_\mathrm{HYD}$ is a decreasing function of distance, and since $L$ is constant,
the electromagnetic part $L_\mathrm{MAG}$ should be an increasing function.
The same should hold for the corresponding parts of the energy to mass flux ratio:
The electromagnetic part $\mu_\mathrm{MAG}$ is proportional to
$L_\mathrm{MAG}$ and thus increses with distance, while the 
hydromagnetic part $\mu_\mathrm{HYD}=\mu-\mu_\mathrm{MAG}$ decreases.
As a result, at least in the region where $V_\varphi$ changes sign, energy 
should be transferred from the matter to the electromagnetic part of the outflow.
This can be realized in a decelerating cold flow, or in an accelerating hot flow
in which part of the enthalpy is transferred to the electromagnetic field.
The last case has been seen in the simulations of the following section, and also in
one of the simulations of \cite{KVKB09}.
Section~5.5 of that paper includes a connection of the possible counter-rotation of the jet
with the poloidal field/streamline shape (which in turn is related to the magnetic energy 
density and the Poynting flux through the bunching function), as well as a
discussion on why previous self-similar solutions of the 
relativistic MHD hot outflow problem \citep{VK03a} do not show counter-rotating jets.

Another connection of counter-rotation with the flow energetics follows from a 
combination of Eqs.~(\ref{L}) and (\ref{mu})
\begin{equation}
\frac{\xi\gamma \varpi\Omega V_\varphi}{\mu c^2} = \frac{L\Omega}{\mu c^2} - \frac{\sigma}{1+\sigma} \,,
  \label{E}
\end{equation}
where $\sigma = \mu_\mathrm{MAG}/ \mu_\mathrm{HYD}$  is the magnetization function. 
Negative $V_\varphi$ requires a sufficiently magnetized flow such that
$\sigma > L\Omega / (\mu c^2 - L\Omega)$.
This in turn implies that $L\Omega $ must be significantly smaller than $\mu c^2$,
contrary to outflows that are strongly Poynting-dominated near their origin
(in which $\mu c^2 \approx \mu_\mathrm{MAG} c^2 = L_\mathrm{MAG}\Omega \approx L \Omega$).
Consequently, the enthalpy part of the energy flux is nonnegligible near the origin of counter-rotating outflows.

\section{Counter-rotation in numerical simulations}

\subsection{Numerical setup}

\begin{figure}
\includegraphics[width=\textwidth]{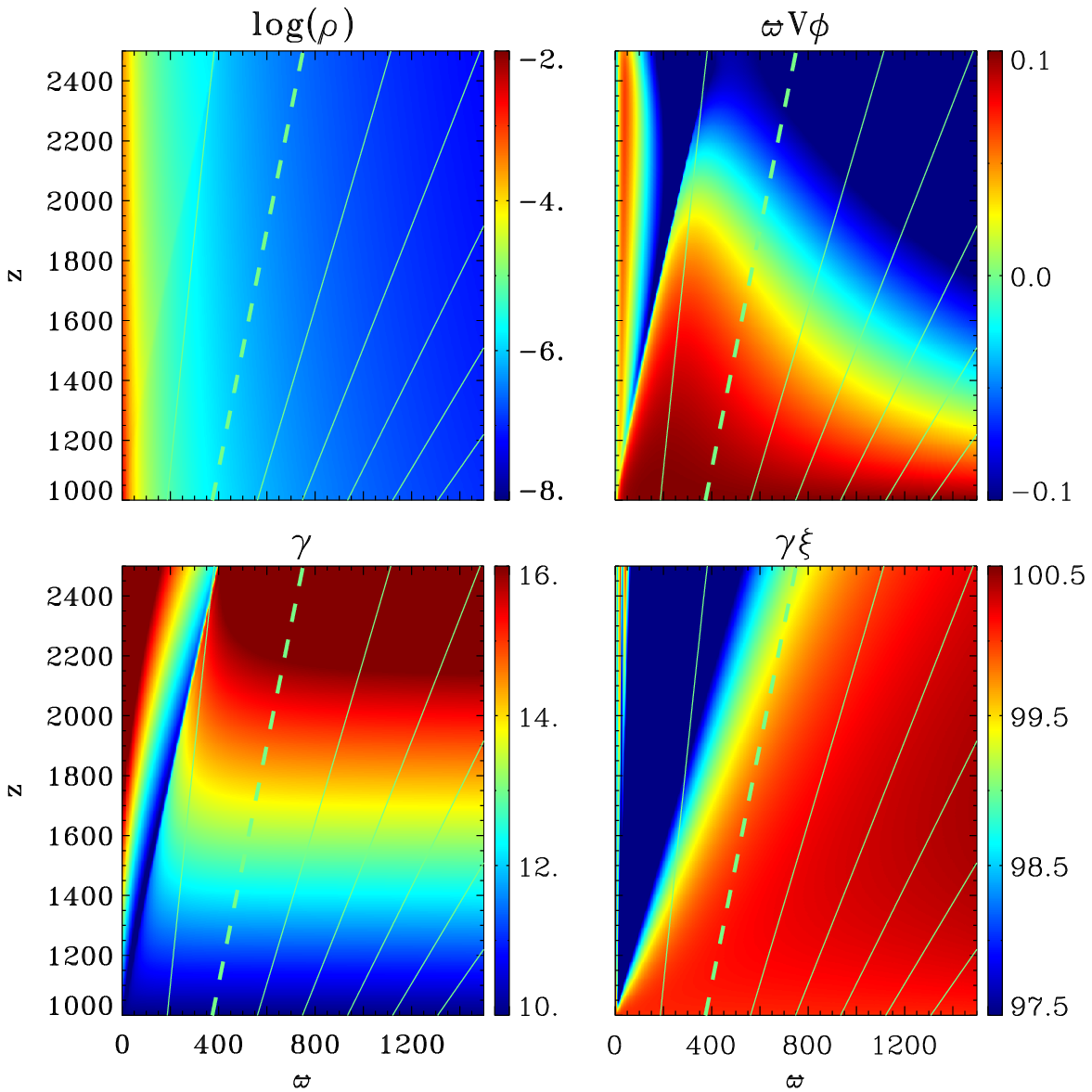}
\caption{Maps of the density (top left), Lorentz factor
  (bottom left), toroidal velocity multiplied by the radius (top right),
  relativistic enthalpy (bottom right), together with the magnetic field lines.
  Whenever the values of each quantity are outside the plotted range, the colors
  saturate to the corresponding min/max value.}
  \label{fig:maps}
\end{figure}

\begin{figure}
\includegraphics[width=\textwidth]{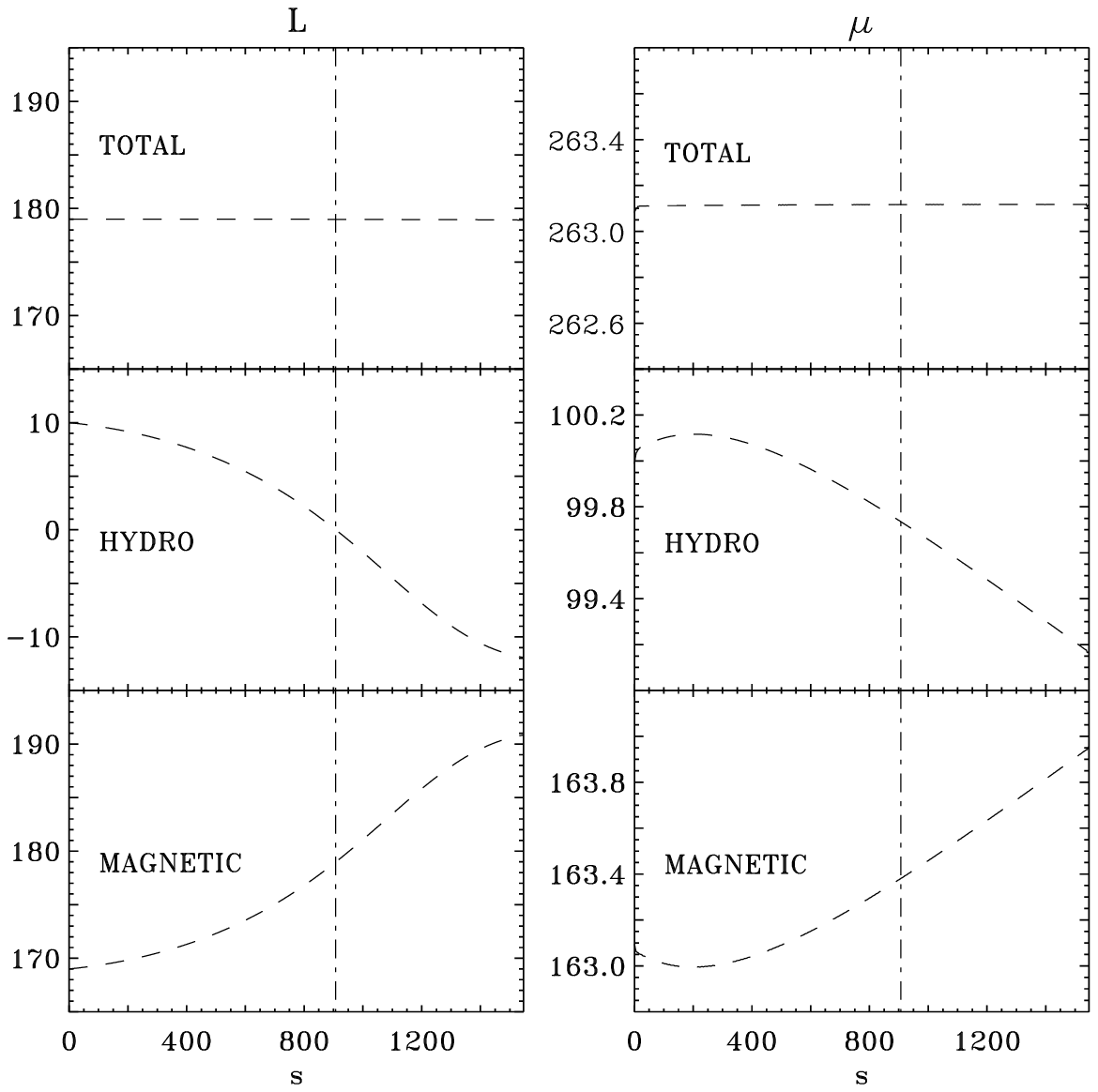}
\caption{The profiles of $L$ (left) and $\mu$ (right) along the dashed
  field line indicated in Fig.~\ref{fig:maps}.
  The total values of the integrals are broken to the hydro and magnetic
  components.
  The vertical dot-dashed line shows the location at which counter-rotation
  starts.}
  \label{fig:graphs}
\end{figure}

In order to illustrate the above demonstration, we perform a numerical
simulation of a jet using the relativistic MHD framework of the PLUTO code
\citep{Mignoneetal07}. We assume axisymmetry, 
an ideal equation of state with polytropic index 4/3,
and focus in the superfast region of the outflow.
The initial conditions provide an equilibrium along the radial direction, but
not in the vertical one.
We choose for the vertical magnetic field a field that decreases along the jet as a power law related to the 
natural transverse expansion of the jet. At a given altitude, close to the axis the vertical magnetic field is 
nearly constant and decreases transversely as $1/\varpi^2$ such that the total magnetic flux does not 
diverge,
\begin{equation}
  B_z = \frac{2c_1c_2}{z^{2/b} + c_2\varpi^2}\,.
\end{equation}
Such a choice, together with a radial magnetic field
\begin{equation}
  B_\varpi = \frac{\varpi}{bz}B_z\,,
\end{equation}
which is consistent with parabolic poloidal field lines $z\propto \varpi^b$ 
\citep{HeyvaertsNorman89},
satisfies the divergence free condition. 

The toroidal magnetic field is deduced from the vertical magnetic field such that the poloidal current 
intensity becomes constant far from the axis. 
The toroidal velocity field is taken to be linear close to the axis consistently with a jet from a solid body 
rotator and decreases at large distances as the inverse of the distance, similarly to the toroidal magnetic 
field,
\begin{equation}
  B_\varphi = -\varpi B_z\,,
\qquad
  V_\varphi = V_{\varphi;0}\frac{c_2\varpi}{z^{2/b} + c_2\varpi^2}\,.
\end{equation}
To start as close as possible from a steady state, we choose the poloidal velocity parallel to the initial poloidal magnetic field (equivalently the toroidal component of the electric field vanishes).
Using the definition of the Lorentz factor we get,
\begin{equation}
  V_\varpi = \sqrt{\frac{1 - 1/\gamma^2 - V_\varphi^2}{1 + B_z^2/B_\varpi^2}}\,, \qquad
  V_z = \frac{B_z}{B_\varpi}V_\varpi\,.
\end{equation}
The equilibrium is statisfied only in the radial direction. We assume a 4/3 polytrope and 
 obtain the initial density and pressure distributions,
\begin{equation}
  \rho_0 = \frac{8c_1^2c_2^2\varpi^2(z^{2/b} - c_2\gamma^2)}{\gamma^4\xi(z^{2/b} + c_2\varpi^2)^3V_
  \varphi^2}\,,
  \qquad
  P = \frac{\xi - 1}{4} \rho_0\,,
\end{equation}
where $\gamma = 10$ and $\xi = 10$ are the initially uniform Lorentz factor and 
specific enthalpy. We choose $c_1 = 10$, $c_2 = 99.5$, $b = 1.5$, and $V_{\varphi;0} = 0.1$ 
for the other constants.
The value of $c_1$ provides the strength of the magnetic field, whereas $c_2$
sets the radial distance beyond which $B_z$ and $V_\varphi$
become pure power laws of $\varpi$.
The exponent $b$ determines the shape of the poloidal field/streamlines.

The size of the box is $\varpi\times z = [0,1]\cdot10^3\times[1,2.5]\cdot10^3$
and is resolved by $512$ zones in each direction.
A stretched grid of $256$ zones extends the radial direction up to
$\varpi = 10000$ in order to minimize boundary effects.
The speed of light is set to unity, the length is normalized in units of the
light cylinder, and the magnetic fields have absorbed the factor
$1/\sqrt{4\pi}$.
Axisymmetric conditions are applied on the left boundary of the computational
domain, outflow on the top and right edges\footnote{We have verified that the
results are robust no matter what boundary conditions are applied on the right
side of the box.} and the initial conditions are kept fixed on the lower
boundary.
We carry out the simulation up to $t = 50000$, but we note that a steady-state
is reached within a fraction of this time.

\subsection{Results}

During the first steps of the simulation the flow reaches a steady-state as the
information from the bottom boundary propagates upwards.
An oblique shock forms close to the axis, a feature which is common to radially
self-similar outflow simulations (see \citealt{MTVMMT08} for a
discussion).
Even though counter-rotation effects take place due to the shock (see
\citealt{Fendt11} for a discussion of the mechanism), 
in this paper we focus farther away from the axis where the flow is smooth.
Figure~\ref{fig:maps} displays the final configuration of the jet at $t = 50000$.
The flow is accelerated along the field lines as it can be seen by the gradual
increase of $\gamma$.
However, the hydrodynamic part of the energy to mass flux ratio, 
$\xi\gamma$, decreases and hence a drop in
$V_\varphi$ is anticipated as shown in Sect.~\ref{sec:energetics}.
Indeed, the toroidal velocity decreases and in fact it becomes negative above a
certain height, beyond which the jet counter-rotates.

In more detail, Fig.~\ref{fig:graphs} shows that the total
angular momentum to mass flux ratio ($L$) and the total energy to mass flux ratio ($\mu$)
are constant along the field/streamlines as expected in axisymmetric steady states.
However, energy is transfered from the matter to the electromagnetic field 
($\mu_\mathrm{HYD}$ decreases and $\mu_\mathrm{MAG}$ increases).
In addition, since the magnetic terms in the energy and angular momentum flux
are proportional, the term $L_\mathrm{HYD}$ is also converted to
$L_\mathrm{MAG}$ in order for the angular momentum flux to remain constant.

Here, we note that the details of the final configuration of the setup depend
weakly on the resolution as well as the location of the outer boundary.
However, the counter-rotation effect we have described is a robust feature that
appears in all variations of our simulated numerical model.
Specifically, the morphology of the integrals and their components have the same
profile in all cases examined, following the prediction of the analytical
demonstration.

\section{Conclusion}

In \cite{Sautyetal12},  as well as in this Letter we have shown that counter-rotation is a signature of the magnetization of a jet, both in 
the classical and the relativistic regimes. Counter-rotation is possible in the following 
 three cases, the first two ones refer to the 
geometry of the flow while the third one is related to the energetics of the jet,
\begin{enumerate}
\item
Gradients of the magnetic field are associated with a compression of the flow
and a sufficiently small $\delta S$ such that Eq.~(\ref{Vphidem3}) is satisfied.
\item
In a smooth flow in which $\delta S \propto \varpi^2$, a decrease of $\mu_\mathrm{HYD}$
may lead to the inequality~(\ref{Vphidem3}). This can happen if the flow is decelerated.
\item
In an ultrarelativistic accelerated or constant flow with $\delta S \propto \varpi^2$, 
part of the decreasing enthalpy flux could be 
transfered to the electromagnetic field, see Sect.~\ref{sec:energetics}. 
Then the hydrodynamic part of the energy to mass flux ratio
$\mu_\mathrm{HYD}$ should be smaller than its value at the Alfv\'en surface to obtain counter-rotation.
\end{enumerate}

In all cases, the role of the magnetic field is critical because it provides the agent 
to absorb the excess of matter angular momentum.
Therefore, counter-rotation is only possible in MHD outflows.
As explained at the end of Sect.~\ref{sec:energetics} the thermal content in these flows is also 
important at least near their origin.

The first case may be obtained if there are surface instabilities at the edge of the jet. In this case although 
the jet expands the flux tubes may squeeze. The changes in the size of the flux tube can induce counter 
rotation.

The second case may occur in FRI jets. As a matter of fact, FRI radio galaxies usually exhibit ultra 
relativistic jets on the parsec scale while the kilo parsec jet is mildly or not relativistic. 
It is usually interpreted as a strong deceleration of the moving plasma  (see \citealp{Melianietal10} and 
references therein). This strong deceleration may 
correspond to the second case where we would observe counter-rotation. This is comparable to the non 
relativistic case studied in \citep{Sautyetal12}. Conversely, if the kilo parsec scale outflow were to be an 
outer component slower than the inner relativistic spine jet that we see on the parsec scale, 
provided that both components are accelerated, then it is likely that the jet rotation sense would be the 
same on all scales.  Measuring the rotation of the flow would thus allow to distinguish 
between a single component decelerating outflow which changes sense of rotation and a two component 
accelerating jet which does not changes rotation. 

The third case corresponds to  a decrease of the total enthalpy budget at large distances. In such 
ultrarelativistic outflows, the transformation of thermal energy flux (enthalpy) to Poynting energy flux would 
induce counter-rotation and a strong increase in the toroidal magnetic field, i.e. matter energy is converted 
into magnetic energy. The change of sign of the rotational speed would induce an increase of 
the Poynting flux and of the magnetic component of the total angular momentum. As a byproduct, the 
toroidal magnetic field would increase as well, creating 
strong gradients of the toroidal magnetic field in the poloidal 
direction. At present, the magnetic field direction is measured via the change of Faraday rotation across 
the jet \citep{Mahmudetal13}.
A complex helicoidal magnetic structure in the knots A and C of M87 jet has already been associated 
with possible counter rotation as we mentioned in the Introduction (\citeauthor{Algabaetal13}, \citeyear{Algabaetal13}, see also \citeauthor{Meyeretal13}, 
\citeyear{Meyeretal13} and \citeauthor{Nakamuraetal10}, 
\citeyear{Nakamuraetal10}).

Altogether then, also in relativistic outflows counter-rotation does not contradict at all magnetic launching.  
Instead, it provides another piece of evidence for the essential role played by the magnetic field in the flow.
It is important to measure the sense of rotation along the jet, because we may get, among others,  an evidence for the 
energetic and angular momentum interchanges in the jet between its fluid and magnetic parts. 

\acknowledgments
The authors thank an anonymous referee for her/his valuable comments which help us to improve the manuscript.
This work has been supported by the
project ``Jets in young stellar objects: what do simulations tell us?'' funded under a
2012 \'{E}gide/France-FCT/Portugal bilateral cooperation Pessoa
Programme and by the Scientific Council of Paris Observatory (PTV program). NV acknowledges the hospitality and support of the Laboratoire Univers et Th\'eories, Observatoire de Paris, during his visit
supported by the Scientific Council of Paris Observatory (PTV program).
TM was supported in part by NASA ATP grant NNX13AH56G.

\end{document}